\documentclass[twoside,a4paper,11pt]{sea10}
\usepackage{graphicx}
\usepackage{hyperref}
\usepackage{movie15}
\usepackage{natbib}
\def\teff{\mbox{$T_{\rm eff}$}}

\def\rv{\mbox{$R_{5495}$}}
\def\al{\mbox{$A(\lambda)$}}
\def\mum1{\mbox{$\mu$m$^{-1}$}}

\begin{document}
\pagenumbering{arabic}
\pagestyle{myheadings}
\thispagestyle{empty}
{\flushleft\includegraphics[width=\textwidth,bb=58 650 590 680]{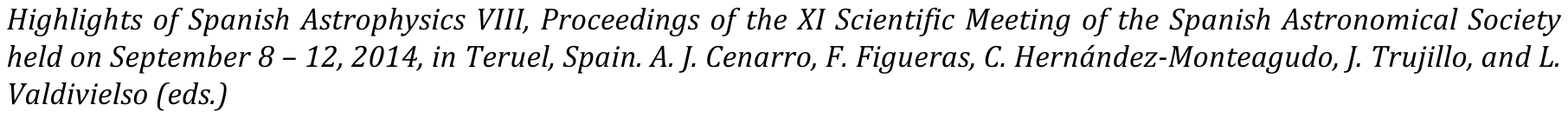}}
\vspace*{0.2cm}
\begin{flushleft}
{\bf {\LARGE
%
Deriving extinction laws with O stars: from the IR to the UV
%
}\\
\vspace*{1cm}
%
J. Ma{\'\i}z Apell{\'a}niz$^1$
%
}\\
\vspace*{0.5cm}
%
$^1$
Centro de Astrobiolog{\'\i}a, INTA-CSIC, Spain\\
%
\end{flushleft}
%
\markboth{
Deriving extinction laws with O stars: from the IR to the UV
}{ 
%
Ma{\'\i}z Apell{\'a}niz
%
}
\thispagestyle{empty}
\vspace*{0.4cm}
\begin{minipage}[l]{0.09\textwidth}
\ 
\end{minipage}
\begin{minipage}[r]{0.9\textwidth}
\vspace{1cm}
\section*{Abstract}{\small
%
We have recently derived a family of extinction laws for 30 Doradus that provides better fits to the optical photometry of obscured stars in 
the Galaxy and the LMC. Simultaneously, we are extending our Galactic O-Star Spectroscopic Survey 
(\href{http://adsabs.harvard.edu/abs/2011hsa6.conf..467M}{GOSSS}) to fainter, more extinguished stars 
to obtain accurate spectral types for massive stars with more than 6 magnitudes of $V$-band extinction. I have combined both lines of research 
with 2MASS, WISE, and Spitzer photometry to obtain the 1-10 micron extinction law for O stars in the solar neighborhood. I present these 
results and compare them with the extinction laws in the same wavelength range derived from late-type stars and H\,{\sc ii} regions. I also
discuss plans to extend the newly derived optical-IR extinction laws to the UV.

%
\normalsize}
\end{minipage}

\section{Introduction}

$\,\!$\indent Walter Baade was once asked if he would become an astronomer if he were born again. He replied that he would only do it if the 
extinction law were the same everywhere. In the mid-twentieth century astronomers worried about the variability of the extinction law in the optical 
but the preoccupation now extends to other ranges, such as the IR and the UV. \citet[CCM]{Cardetal89} produced the first single-parameter family of
extinction laws that extended from the IR to the UV. In \citet{Maiz13b} we described some of the problems with the CCM family laws and in
\citet{Maizetal14a} we proposed an alternative family of extinction laws for the optical range that used the CCM results for the two other ranges. In
this contribution we proceed from the IR to the UV by [a] extending our analysis to longer wavelengths using Galactic O stars, [b] summarizing the new
laws of \citet{Maizetal14a} in the optical, and [c] presenting our plans to produce a new family of extinction laws that also reaches the UV. The new
analysis relies heavily on the Galactic O-Star Spectroscopic Survey (GOSSS), described in \citet{Maizetal11}. To date, we have published the
blue-violet spectra and the spectral types of 448 Galactic O stars \citep{Sotaetal11a,Sotaetal14} and we have already observed over 2000 stars (of O
type and others). See the contribution by Ma{\'\i}z Apell\'aniz et al. in these proceedings for additional GOSSS results on a new spectral library, a
spectral classification tool, and a first list of stars wrongly classified as O in the literature.

Before proceeding, it is worth reiterating something we have mentioned elsewhere: $R_V$, the band-integrated ratio of total to selective extinction 
depends not only on the type of dust but also on its amount and on the type of stellar SED \citep{Maiz13b}. Therefore, {\it $R_V$ should not be used 
to characterize an extinction law}. Instead, one should use monochromatic quantities such as \rv.

\begin{figure}
\centerline{
\includegraphics[width=0.550\linewidth, bb=38 38 556 556]{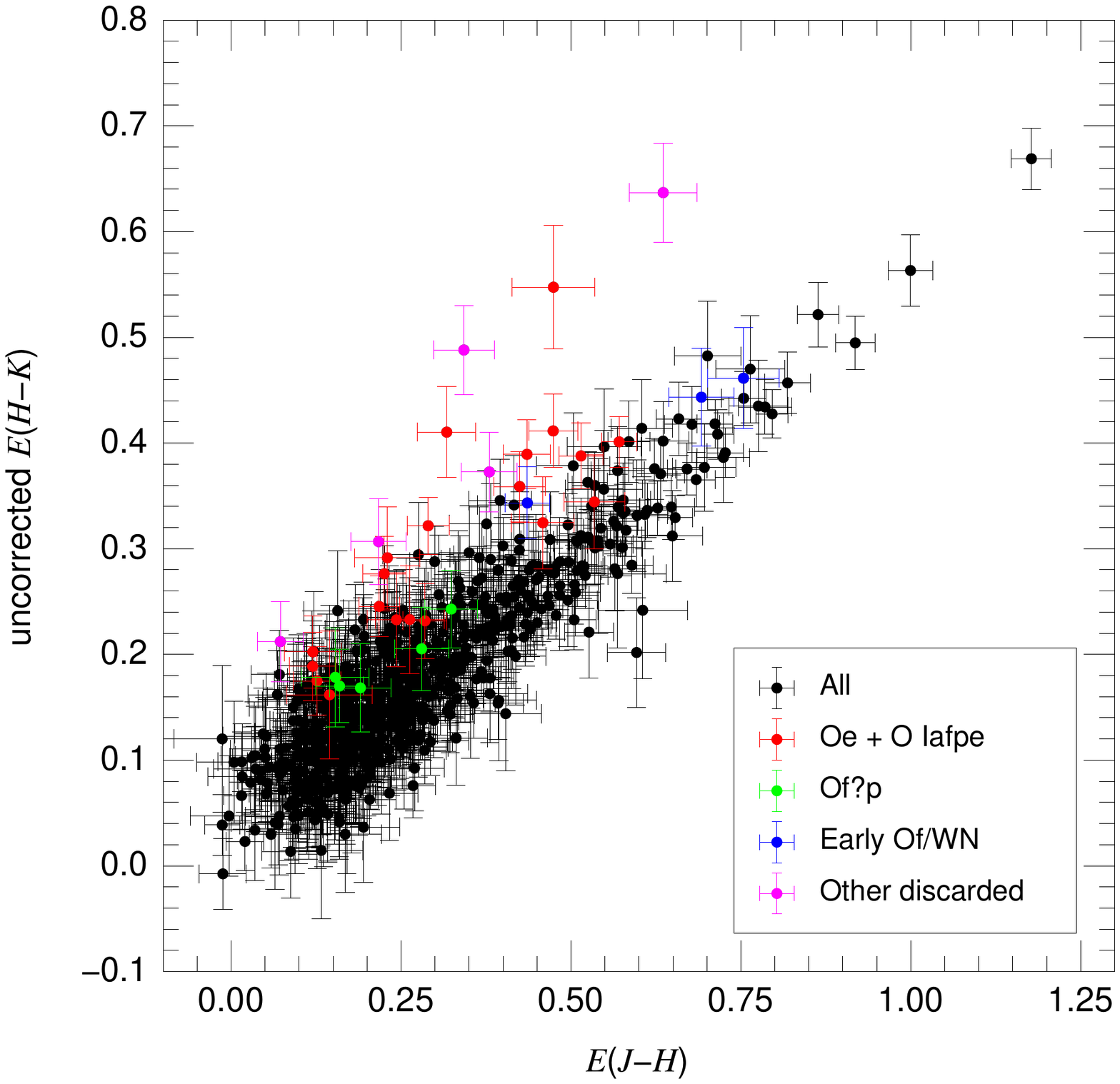}   
}
\centerline{
\includegraphics[width=0.550\linewidth, bb=38 38 556 556]{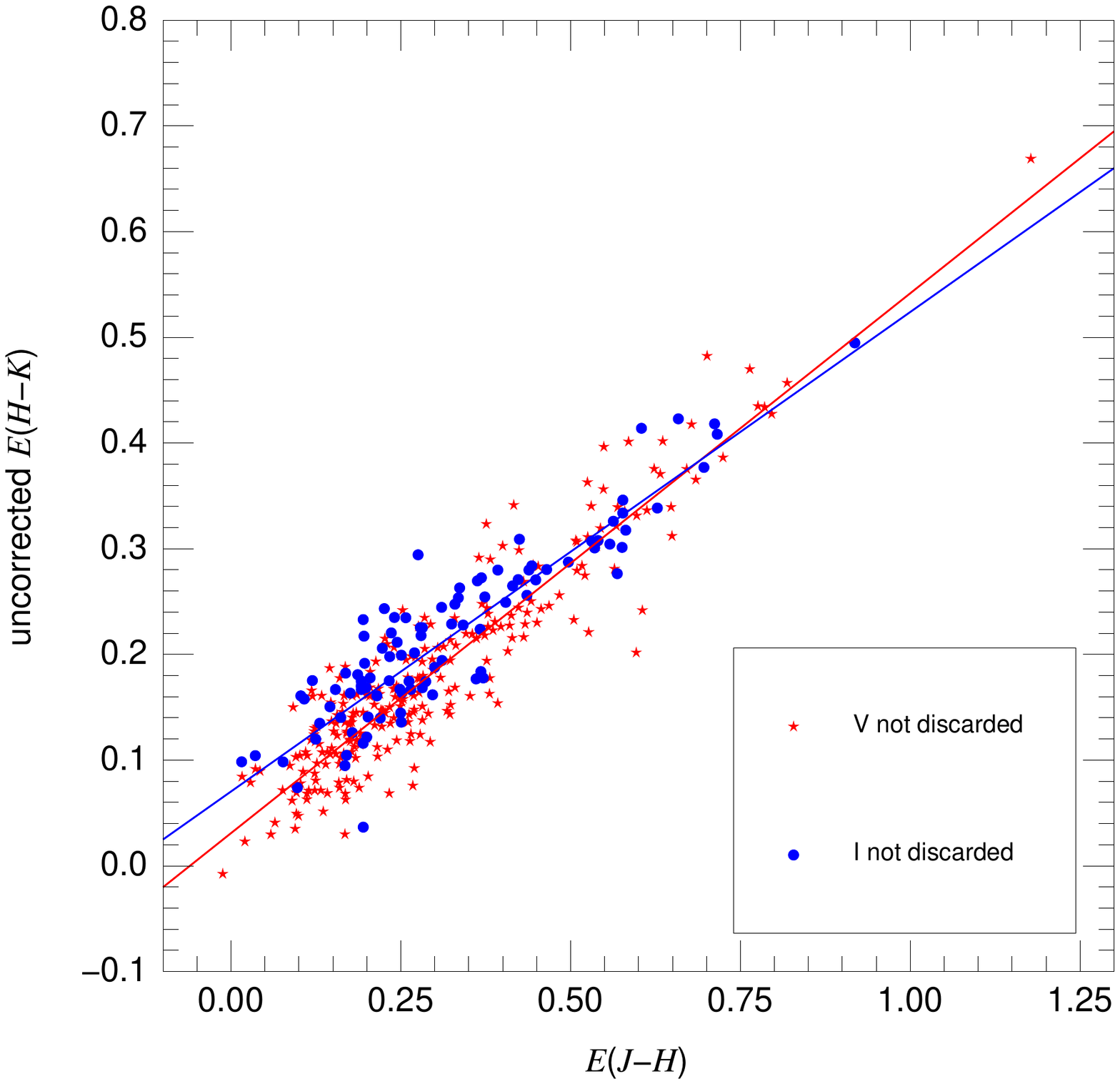} ~ 
\includegraphics[width=0.550\linewidth, bb=38 38 556 556]{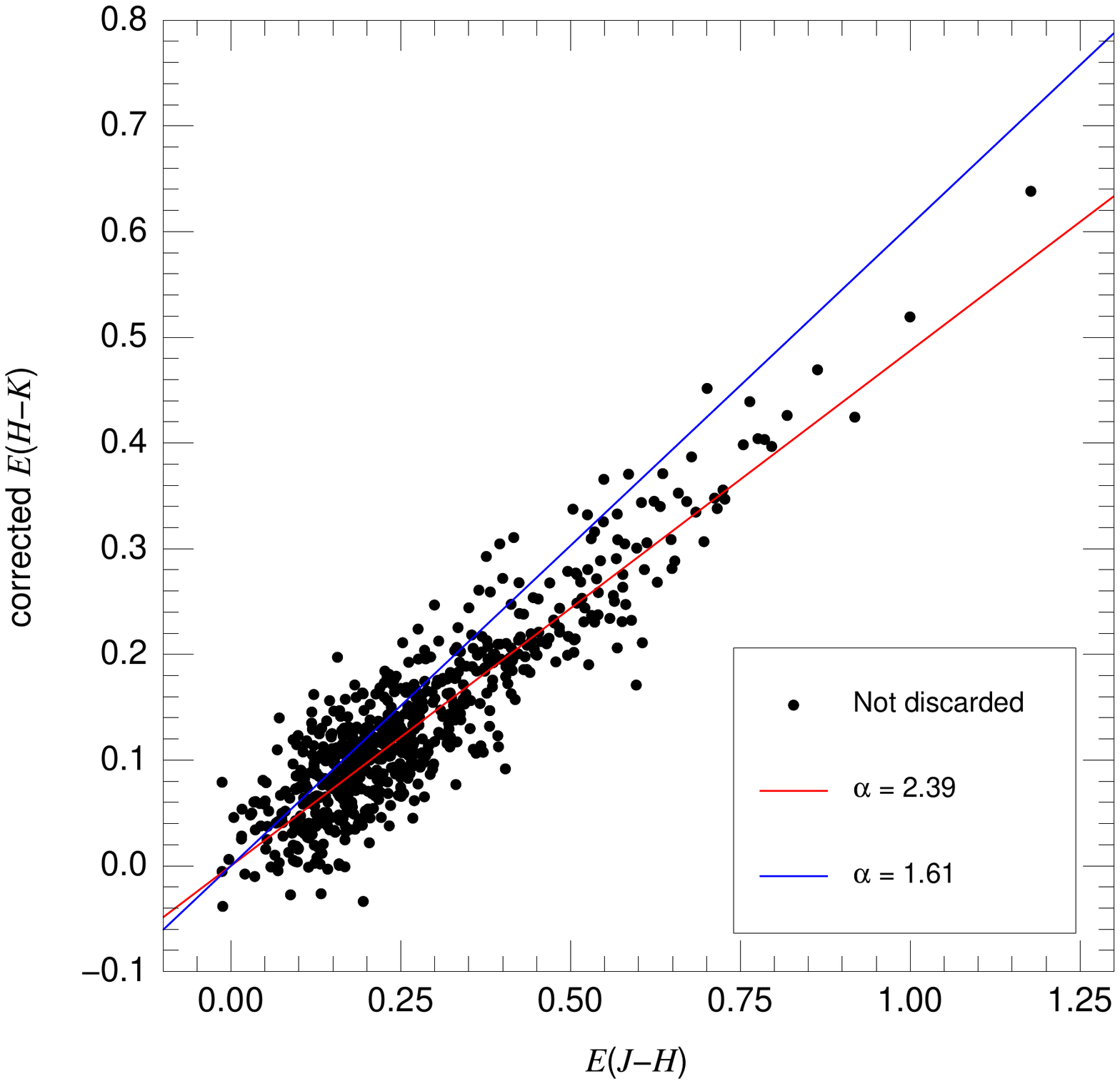}
}
\caption{(top) Uncorrected $E(H-K_{\rm s})$ vs. $E(J-H)$ plot for the GOSSS O-star sample. (bottom left) Same plot (without error bars) for spectral
luminosity classes I and V after discarding stars with likely IR excesses. The linear fits to the two samples used to determine the wind correction to
$E(H-K)$ are also shown. (bottom right) Final plot after applying the luminosity-class-dependent corrections. The expected extinction effect for two 
power laws with different exponents is also shown.}
\label{fig1}
\end{figure}

\section{The IR}

$\,\!$\indent In the IR, the CCM laws use a power law $\al/A(1\,\mu{\rm m}) = x^{\alpha}$ ($x\equiv 1/\lambda$, with $\lambda$ in $\mu$m) with
$\alpha = 1.61$ fixed, which was adopted from
\citet{RiekLebo85}. More recent works in the IR have found different values of the exponent, possible variability from one sightline to another, and
complex behaviors as a function of wavelength \citep[e.g.][]{Mooretal05,Romaetal07,Nishetal09,Gaoetal09,Schoetal10}. Those papers, however, use as 
reference targets cool stars and/or H\,{\sc ii} regions, whose intrinsic SEDs are subject to larger uncertainties than those of OB stars. Furthermore, 
if one uses different objects and/or environments to derive extinction laws, one cannot be certain that possible differences are intrinsic to the 
extinction law since they may be instead caused by the sample heterogeneity. Therefore, if one derives extinction laws in different wavelength regimes 
(e.g. UV, optical, and IR), one should use at least a subsample that spans all of them. This is what \citet{FitzMass09} attempted with OB stars but 
their sample was small (14 objects) and only included targets with $A_V$ up to $\sim$4, for which extinction in the IR is small, leading to a 
measurement of the extinction law with relatively large uncertainties.

I have started reanalyzing the IR extinction laws with GOSSS, taking advantage of:

\begin{itemize}
 \item Its large sample size of Galactic targets with spectral types: we are getting close to 1000 O and 1000 B stars observed.
 \item The accuracy and detail of its spectral types, which allows for the detection and elimination of peculiar objects that may have non-standard
        intrinsic SEDs in the IR (e.g. strong winds, circumstellar disks, obscured companions). 
 \item Its large range of extinctions (up to $A_V \sim 9$) that includes newly obtained high-extinction objects observed with large telescopes. 
 \item The recent availability of WISE and Spitzer photometry for many of the GOSSS targets. 
\end{itemize}

I have proceeded in the following way:

\begin{enumerate}
 \item I collected the 2MASS $J+H+K_{\rm s}$, WISE $W1+W2+W3$, and Spitzer IRAC $3.6+4.5+5.8+8.0$ photometry for all the O stars in GOSSS as of early
       2014. The WISE bright stars were corrected of photometric bias\footnote{See
       \url{http://wise2.ipac.caltech.edu/docs/release/allsky/expsup/sec6_3c.html}.}.
 \item I assigned TLUSTY intrinsic SEDs \citep{LanzHube03} to each star according to the spectral type-\teff\ calibration of \citet{Martetal05a}
       adapted to the new spectral type subdivision \citep{Sotaetal14} using the grid of \citet{Maiz13a}. Note that TLUSTY does not include wind
       effects, so small corrections are required in the IR, and that those corrections are expected to be larger for higher luminosity stars.
 \item I calculated the color excesses $E(J-H)$ and $E(H-X)$ (where $X$ is any of the 2MASS + WISE + Spitzer IRAC filters above except $J$ or $H$)
       for each star, discarded the stars with likely IR excesses (Figure~\ref{fig1}, top), divided the sample by spectral luminosity classes 
       (I to V), and fitted straight lines to the $E(H-X)$ vs. $E(J-H)$ for each subsample (Figure~\ref{fig1}, bottom left). The intercepts were then
       used to produce luminosity-class-dependent corrections for $E(H-X)$ and corrected plots (Figure~\ref{fig1}, bottom right). Only the 
       case where $X=K_{\rm s}$ is shown in Figure~\ref{fig1} but the process was repeated for all filters. For $K_s$ the correction increases 
       monotonically from 0.031 mag (luminosity class V) to 0.070 magnitudes (luminosity class I) and the behavior is simlar for other filters. 

\begin{figure}
\centerline{
\includegraphics[width=0.550\linewidth, bb=38 38 556 556]{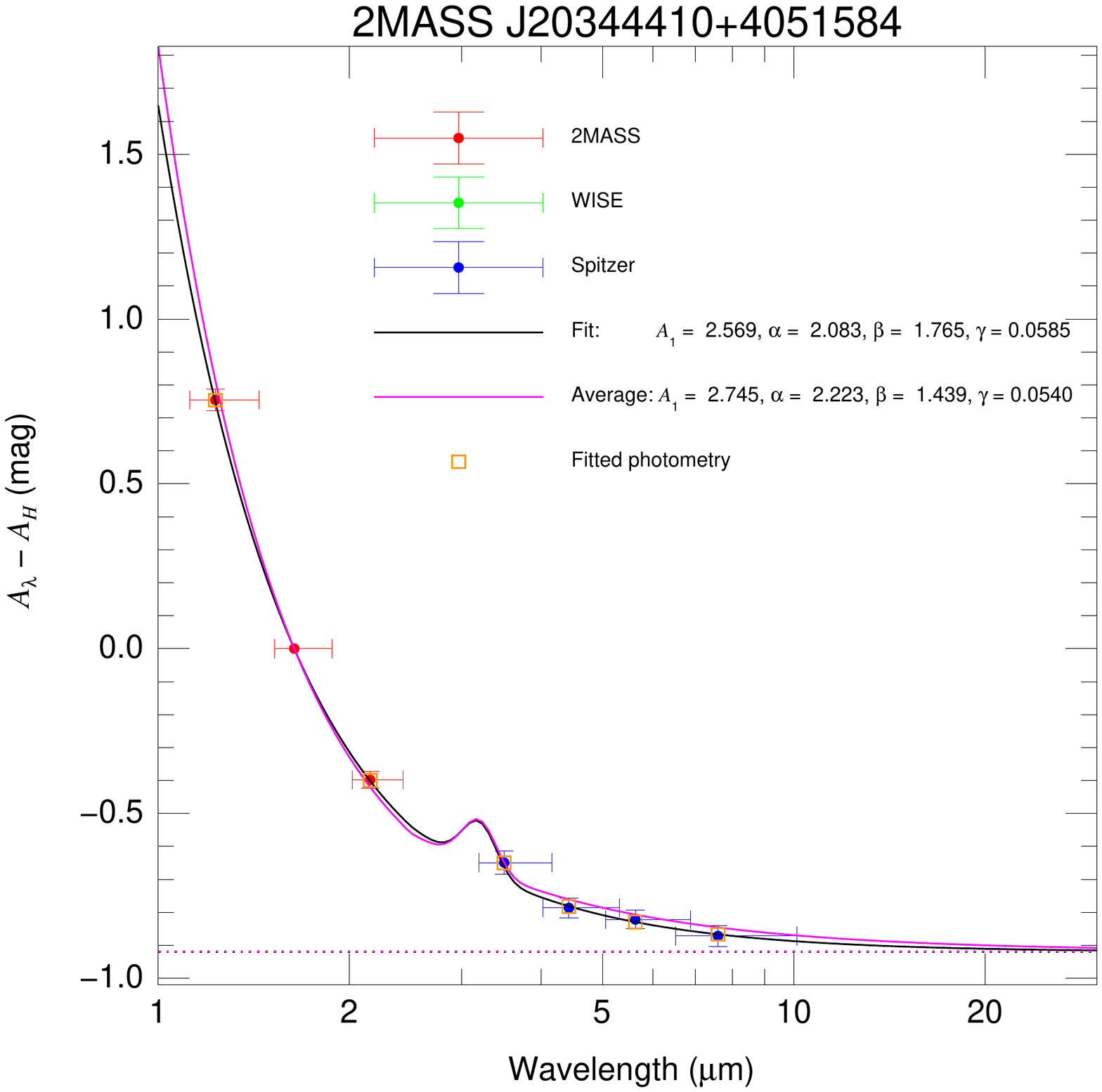} ~ 
\includegraphics[width=0.550\linewidth, bb=38 38 556 556]{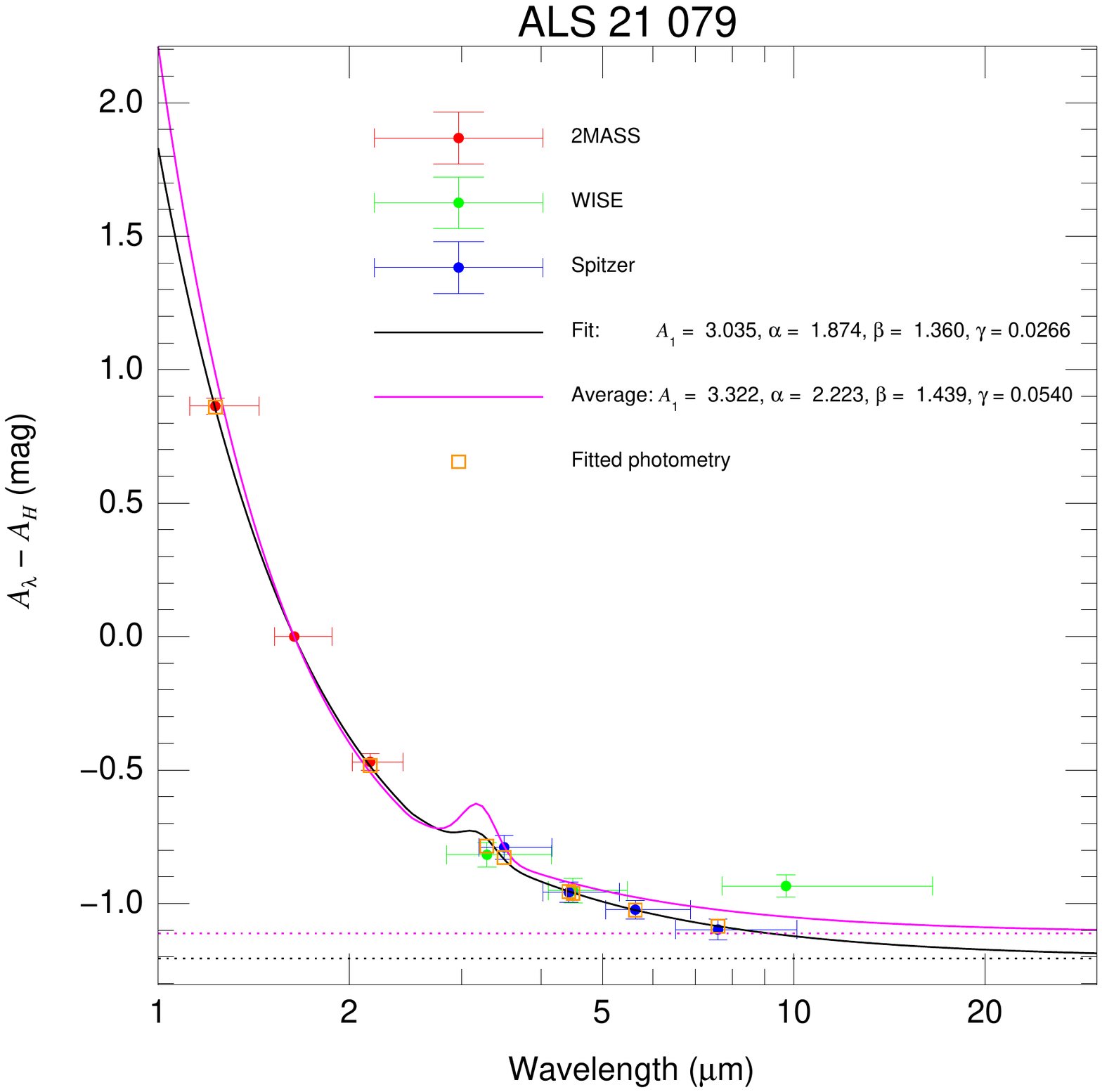}
}
\centerline{
\includegraphics[width=0.550\linewidth, bb=38 38 556 576]{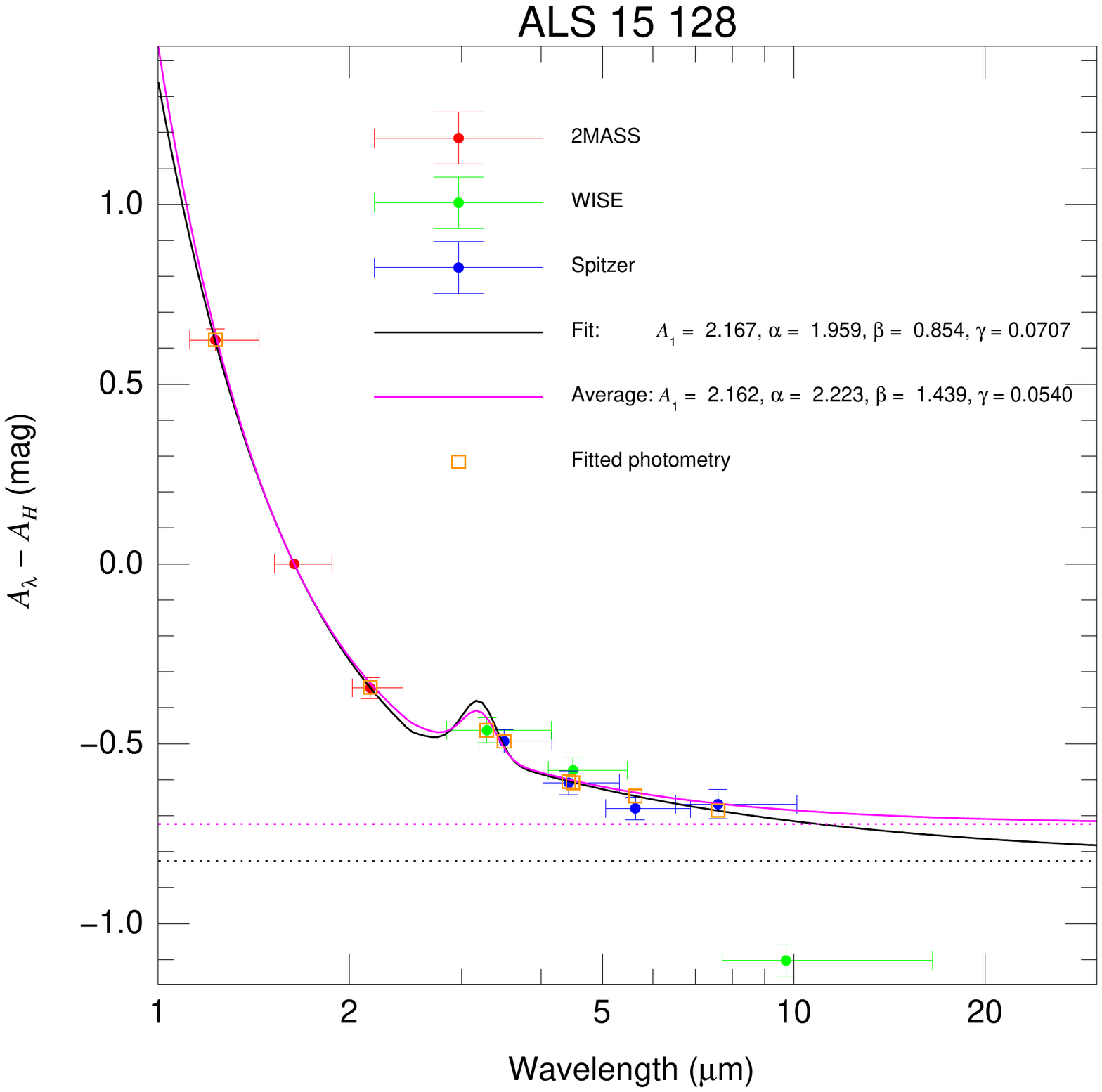} ~ 
\includegraphics[width=0.550\linewidth, bb=38 38 556 576]{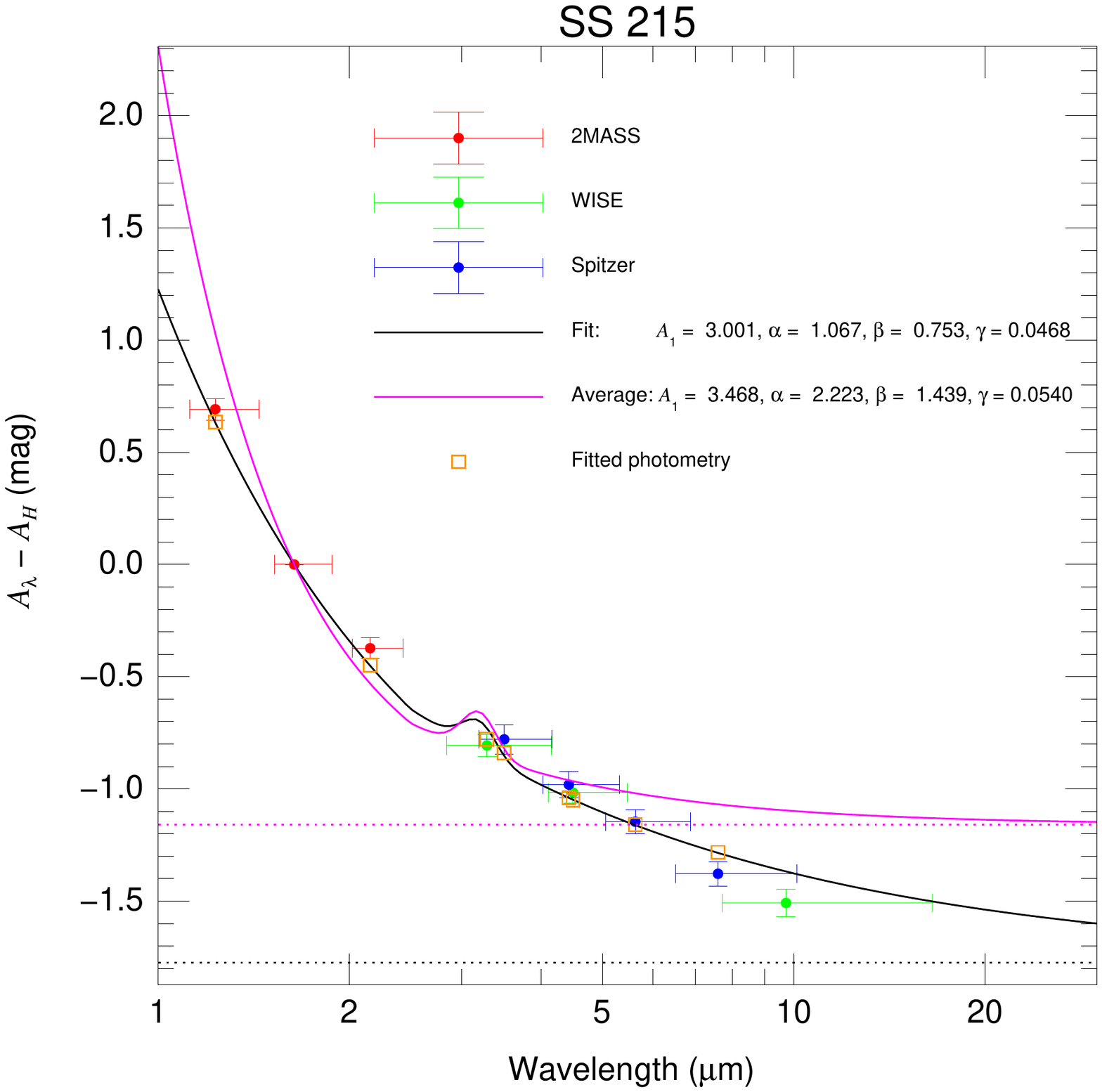}
}
\caption{IR extinction-law plots for four GOSSS stars. 2MASS~J20344410+4051584 is a case with only 2MASS + Spitzer IRAC photometry. ALS~21\,079 shows 
the possible effect of the 10~$\mu$m silicate band in W3. ALS~15\,128 apparently has a 10~$\mu$m excess. SS~215 is a case where strong winds 
invalidate our simplified approach to calculate the IR extinction law.}
\label{fig2}
\end{figure}

 \item The next effect was to select a extinction law model. In the NIR a power law is commonly used but beyond 2.5~$\mu$m different structures are
       expected \citep{Fritetal11}. I chose a simplified three-parameter ($\alpha+\beta+\gamma$) model for $A_\lambda/A_1$ ($A_1$ being the 
       extinction at 1 $\mu$m) consisting of:
 \begin{itemize}
  \item A power law with exponent $\alpha$ between 1~$\mu$m and 2.5~$\mu$m.
  \item Another power law with exponent $\beta$ beyond 2.5~$\mu$m. The two power laws are joined ``a la Kroupa'' at 2.5~$\mu$m but in order to 
        preserve the derivability a weighted sum of the two power laws is applied between 2.4~$\mu$m and 2.6~$\mu$m. 
  \item A gaussian component centered at 3.3~$\mu$m with a $\sigma$ of 0.2~$\mu$m and a peak intensity of $\gamma$. The purpose of this component is
        to model the combined effect of H$_2$O and aliphates at these wavelengths, which are the lines with the stronger expected equivalent 
        widths in the 1-8 $\mu$m range \citep{Fritetal11}.
 \end{itemize}
 \item I fitted the model to the most extinguished stars in the GOSSS sample using a $\chi^2$-minimizing code with four free parameters:
       $\alpha$, $\beta$, $\gamma$, and $A_1$. Note that different stars may have different photometric bands available:
       we only used those cases with at least five existent magnitudes. Also note that WISE W3 was not used in the fitting for three reasons: [a] low
       S/N and confusion with background, [b] possible presence of silicate absorption around 10~$\mu$m \citep{Fritetal11}, and [c] possible presence
       of excesses due to circumstellar cool material. Nevertheless, the W3 band photometry is shown in the plots in Figure~\ref{fig2} for reference.
 \item An average extinction law was built from a selection of the fitted stars. The average values found were: $\alpha = 2.162$, $\beta = 1.439$, and
       $\gamma = 0.00540$. The extinction in the NIR is clearly steeper than the CCM value and then flattens beyond 2.5 $\mu$m, in agreement with 
       other recent results. The intensity of the H$_2$O + aliphates peak around 3.3~$\mu$m is roughly consistent with the extinction law of 
       \citet{Fritetal11}.
 \item The four examples in Figure~\ref{fig2} can be used to give an idea of the diversity of the data. SS 215 is an O2 If*/WN5 star
       \citep{Sotaetal14} that shows the largest apparent deviation from the average extinction law. However, such stars have strong winds whose
       effect in the IR SED is likely not to be corrected by the simplified procedure described here. Therefore, the low value of $\alpha$ measured 
       is likely to be an artifact and not a real extinction effect. 
\end{enumerate}

The results presented here are preliminary. We are currently working on expanding our high-extinction sample before publishing our final rsults on the
IR extinction law using GOSSS stars.

\section{The optical}

$\,\!$\indent \citet{Maizetal14a} have combined the spectral types from the VLT-FLAMES Tarantula Survey \citep[VFTS][]{Evanetal11a,Walbetal14} with 
photometry from the HST/WFC3 Early Release Science (GO 11\,360) to produce a new family of extinction laws for the optical region. The new laws:

\begin{itemize}
 \item Maintain the overall behavior and good characteristics of CCM: they are a single-parameter (\rv) continuous and differentiable family.
 \item Eliminate the $U$ band excesses detected in CCM for all values of \rv.
 \item Alleviate the wiggles introduced by the use of the seventh degree polynomial in $1/\lambda$ used by CCM and make the extinction laws more 
       Whitford-like \citep{Whit58,ArdeVird82}.
\end{itemize}

The new family of extinction laws produce significant better fits for the optical-NIR photometry of both Galactic and 30 Doradus 
targets\footnote{The differences between MW, LMC, and SMC extinction laws studied in the past refer to the UV.}. However, they should not be the final
word for four reasons:

\begin{itemize}
 \item \citet{Maizetal14a} assumed the \citet[the same as CCM]{RiekLebo85} exponent in the IR, which is too shallow, as we have seen here. Therefore, 
       the optical laws should be ``stitched'' to a more correct IR law (or laws, if it is finally shown that there is real variation in the IR).
 \item The more correct way of deriving extinction laws is with spectrophotometry, not with photometry, thus avoiding interpolation in wavelength. 
       This is easier to do in the optical than in the IR and we have already obtained data with this purpose.
 \item The ``stitching'' with the UV should also be revisited (see next section).
 \item Finally, some discrete absorption features in the ISM have been extensively studied in the optical and some of them are highly correlated with 
       extinction. Therefore, it should be possible to include them in a high-spectral-resolution version of the extinction law, a goal towards which
       we are also working in \citep{Maizetal14b}.
\end{itemize}

In summary, once a new IR extinction law is obtained, it should be possible to improve upon the optical results of \citet{Maizetal14a}.

\section{Onto the UV}

$\,\!$\indent The non-specialist astronomer may think that the UV extinction laws were long ago settled with IUE. However, that is far for the truth, 
as the following points show:

\begin{itemize}
 \item The CCM laws used a single parameter ($R_V$ or, more properly, \rv) to describe the whole UV-to-IR wavelength range. However,
       \citet{FitzMass07} claim that with the exception of a few curves with large values of \rv, the UV and IR portions of Galactic extinction
       curves are not correlated with each other, which is in direct contradiction with CCM.
 \item \citet{FitzMass07} also find that ``the central position of the 2175 \AA\ extinction bump is mildly variable, its width is highly variable, 
       and the two variations are unrelated.'' 
 \item It is usually expressed that for the SMC there is no 2175 \AA\ bump in the extinction law. Actually, \citet{Gordetal03} find four SMC 
       sightlines without a bump and one with a weak bump. More recently, \citet{MaizRubi12} studied four additional SMC sightlines and found one 
       with a significant bump, two with a weak bump, and one without it. Clearly, more sightlines are needed (an ubiquitous issue with UV
       extinction).
 \item A bump-less extinction curve may not be exclusive to low metallicity environments such as the SMC: \citet{Valeetal03} found an example in
       the Milky Way.
 \item In some objects geometry may be at work. As shown in another contribution to these proceedings by Ma{\'\i}z Apell{\'a}niz et al., a large 
       fraction of the UV light coming from Herschel 36 and its surroundings is actually scattered radiation from the H\,{\sc ii} region, an effect
       that is also known to be important in the Orion Nebula. For Herschel 36, IUE was capable of resolving the star from the nebulosity but 
       in more distant objects we may be considering the joint flux instead of just that received directly from the star.
 \item Speaking of geometry, when the UV radiation originates from multiple sources or scattered radiation is included, one should use the term
       ``attenuation law'' instead of ``extinction law''. A popular example is the Calzetti law \citep{Calz01}. 
\end{itemize}

In summary, there is a relatively large degree of confusion with UV extinction. We know it is highly variable but there is contradictory information
in the literature regarding how that variation takes place. This hampers the solution of the ultimate questions of the origin of the extinction law
and its dependence on metallicity and environment. Fortunately, there are two lines of work that may help us clear the waters.

On the one hand, IUE has an excellent archive that can be combined with new data and techniques. Modern surveys provide optical-IR photometry of 
better quality than what was available until recently. Spectroscopic surveys such as GOSSS and new spectral libraries and modelling can also constrain 
the intrinsic SEDs of the sources better and reduce systematic errors. A reanalysis of this combination of old and new data is a necessary step that
we plan to attack in the near future.

On the other hand, new UV spectroscopy is clearly required. The number of IUE spectra of Galactic sightlines with large \rv\ is very small. The number
of studied extragalactic sightlines (even for the MCs) is also small and, in many cases, with very low extinctions (which amplifies the effect of 
systematic errors). Along this line, it would be very useful to obtain UV spectroscopy of some of the sources in \citet{Maizetal14a}, since we would
kill those two birds (large \rv\ and the LMC) with one stone.

%
%
\small  
%
\section*{Acknowledgments}   
$\,\!$\indent I acknowledge support from [a] the Spanish Government Ministerio de Educaci\'on y Ciencia through grants AYA2010-15081 and AYA2010-17631
and [b] the Consejer{\'\i}a de Educaci{\'o}n of the Junta de Andaluc{\'\i}a through grant P08-TIC-4075.

%
%
%
\bibliographystyle{aj}
\small
\bibliography{general}

\end{document}